\documentstyle[12pt]{article}
\begin{document}
\begin{center}
\textbf{
CASIMIR THEORY OF THE RELATIVISTIC PIECEWISE UNIFORM STRING}\\
\bigskip
I. BREVIK\\ 
Division of Applied Mechanics,\\
 Norwegian University of Science and Technology,\\
N-7034 Trondheim, Norway\\
E-mail: iver.h.brevik@mtf.ntnu.no
\end{center}
\begin{abstract}
The Casimir energy for the transverse oscillations of a piecewise uniform closed string
is calculated. The string is relativistic in the sense that the velocity of transverse
waves is always equal to $c$. The great adaptibility of this string model with respect 
to various regularization methods is pointed out. We survey several regularization methods:
the cutoff method, the complex contour integration method, and the zeta-function method. The
most powerful method in the present case is the contour integration method. The Casimir
energy turns out to be negative, and more so the larger is the number of pieces in the
string. The thermodynamic free energy $F$ is calculated for a two-piece string in the limit
when the tension ratio $x=T_{I}/T_{II}$ approaches zero. For large values of the length
ratio $s=L_{II}/L_I$, the Hagedorn temperature becomes proportional to $\sqrt{s}$.
\end{abstract}

\section{Introduction}
In the standard theory of closed strings - whatever the string is taken to be in Minkowski space
or in superspace - one usually assumes that the string is {\em homogeneous}, i.e. that the tension $T$ is the same everywhere. The {\em composite} string model, in which the string is assumed to consist of two or more 
separately uniform pieces, is a variant of the conventional theory. The system is relativistic, in the sense that the velocity $v_s$  of transverse sound is in each of the pieces assumed to be equal to the velocity of 
light:
\begin{equation}
v_s=\sqrt{T/\rho}=c.
\label{1}
\end{equation} 
Here $T$ and $\rho$ (the density) refer to the piece under consideration. At each junction between pieces of different material there are two boundary conditions: the transverse displacement $\psi = \psi(\sigma,\tau)$ itself, as well as the transverse force $T\partial \psi/\partial \sigma$, must be continuous. Combining Eq.(1) with the wave equation
\begin{equation}
(\frac{\partial^2}{\partial\sigma^2}-\frac{\partial^2}{\partial\tau^2})\psi=0,
\label{2}
\end{equation}
one can calculate the eigenvalue spectrum and the Casimir energy of the string.

The composite string model was introduced in 1990 \cite{brevniels90}; the string was there assumed to consist of two pieces $L_I$ and $L_{II}$. The dispersion equation was derived, and the Casimir energy calculated for various integer values of the length ratio $s=L_{II}/L_I$. Later on, the composite string model has been generalized and studied from various points of view [2-10]; we may mention, for instance, that the recent paper of Lu and Huang [9] discusses the Casimir energy for a composite Green - Schwarz superstring.

Some reasons why the composite string model turns out to be an attractive model to study are the following. First, if one performs Casimir energy calculations, one finds that the system is remarkably easy to regularize: one has access to the cutoff method [1], the complex contour integration method [3-5, 7], or the Hurwitz $\zeta-$ function method [2, 4, 5, 7] ( [8] contains a review of the various regularization methods). As a physical result of the Casimir energy calculations it is also worth noticing that the energy is in general nonpositive, and is more negative the larger the number of uniform pieces in the string is.

The composite string model may moreover serve as a useful two-dimensional field theoretical model in general. The hope is that such a model can help us to understand the issue of the energy of the vacuum state in two-dimensional quantum field theories, what is quite a compelling goal. As a peculiar application, perhaps can this particular string model even play a role in the theories of the early universe. The notable point is here that the string can in principle adjust its zero point energy: the energy always becomes diminished if the string divides itself into a larger number of pieces.

It is also to be noted that there are strong formal similarities between this kind of theory and the phenomenological electromagnetic theory in material media satisfying the condition $\varepsilon \mu =1$, $\varepsilon$ denoting the permittivity and $\mu$ the permeability of the medium [11]. Obviously, the basic reason why the two theories become so similar is that the relativistic invariance is satisfied in both cases; in the string case through Eq.(1), in the electromagnetic case through the equation $\varepsilon \mu=1$.

In the following we put $ \hbar=c=1$.

\section{Two-Piece String}
\subsection{Dispersion Relation}
Let the two junction points, lying at $\sigma=0$ and $\sigma=L_I$, separate the type $I$ and type $II$ pieces from each other. The total length of the closed string is $L=L_I+L_{II}$. We define $x$ to be the tension ratio and define also the function $F(x)$:
\begin{equation}
x=\frac{T_I}{T_{II}},~~~~~F(x)=\frac{4x}{(1-x)^2}.
\label{3}
\end{equation}
The dispersion equation becomes
\begin{equation}
F(x)\sin^2(\frac{\omega L}{2})+\sin \omega L_I\sin \omega L_{II}=0.
\label{4}
\end{equation}
The Casimir energy $E$ of the system is defined as the zero-point energy $E_{I+II}$ of the two parts, minus the zero-point energy of the uniform string:
\begin{equation}
E=E_{I+II}-E_{uniform}=\frac{1}{2}\sum \omega_n-E_{uniform}.
\label{5}
\end{equation}
Here the sum goes over all eigenstates, with account of their degeneracy. It is irrelevant whether $E_{uniform}$ is calculated for type $I$ material or type {II} material in the string, the reason for this being the relativistic invariance expressed by Eq.(1).

We will consider three different methods for regularizing the Casimir energy. 
\subsection{Cutoff Regularization}
The simplest way to proceed [1] is to introduce a function $f=\exp (-\alpha \omega_n)$, with $\alpha$ a small positive parameter, and to multiply the nonregularized expression for $E$ by $f$ before summing over the modes.

We consider first the case of a {\it uniform} string, corresponding to $x=1$. The dispersion equation (4) yields the eigenvalue spectrum $\omega L=1$, which means
\begin{equation}
\omega_n=2 \pi n/L,~~~~n=1,2,3,...
\label{6}
\end{equation}
Taking into account that these modes are degenerate, we find for the zero-point energy
\begin{equation}
E_{uniform}=\frac{L}{2\pi\alpha^2}-\frac{\pi}{6L}+O(\alpha^2).
\label{7}
\end{equation}
Let us next consider the limiting case $x \rightarrow 0$ (we let $T_I\rightarrow 0$ while keeping $T_{II}$ finite). The dispersion relation allows two sequences of modes,
\begin{equation}
\omega_n=\pi n/L_I,~~~~~~\omega_n=\pi n/L_{II},~~~~n=1,2,3,...
\label{8}
\end{equation}
If $s$ denotes the length ratio,
\begin{equation}
s=L_{II}/L_I,
\label{9}
\end{equation}
we then get the simple formula for the Casimir energy
\begin{equation}
E=-\frac{\pi}{24 L}(s+\frac{1}{s}-2).
\label{10}
\end{equation}
Now let $s$ be an {\it odd} integer. The dispersion equation yields one degenerate branch, determined by 
\begin{equation}
\sin\omega L_I=0,~~~~\omega L_I=\pi n,
\label{11}
\end{equation}
and there are in addition $\frac{1}{2}(s-1)$ nondegenerate double branches, determined by solving an algebraic equation of degree $\frac{1}{2}(s-1)$ in   $\sin^2\omega L_I$. The frequency spectrum can be expressed as 
\begin{equation}
\omega L_I=\left\{ \begin{array}{ll}
                    \pi (n+\beta),\\
                     \pi (n+1-\beta),
                     \end{array}
            \right.
\label{12}
\end{equation}
where $n=0,1,2,...$, and where $\beta$ is a number in the interval $0< \beta \leq \frac{1}{2}$. Each double branch yields the four solutions $\pi \beta$, $\pi (1-\beta)$, $\pi(1+\beta)$, and $\pi(2-\beta)$ for $\omega L_I$ in the region between $0$ and $2\pi$. 

Introducing for convenience the abbreviation $t=\pi\alpha(s+1)/L$, we obtain
\begin{equation}
E({\rm degenerate~~ branch})=\frac{1}{\alpha t}-\frac{t}{12\alpha}+O(t^2),
\label{13}
\end{equation}
\begin{equation}
E({\rm double~~ branch})=\frac{1}{\alpha t}+\frac{t}{6\alpha}-\frac{t}{4\alpha}[\beta^2
+(1-\beta)^2]+ O(t^2).
\label{14}
\end{equation}
We replace $\beta$ by $\beta_i$, sum (14) over all $\frac{1}{2}(s-1)$ double branches, and add (13) to obtain $E_{I+II}$. Subtracting off the uniform string result (7), and letting $t \rightarrow 0$, we get the Casimir energy for odd $s$, 
\begin{equation}
E=\frac{\pi s(s-1)}{12L}-\frac{\pi (s+1)}{4L}\sum_{i=1}^{(s-1)/2}[\beta_i^2+(1-\beta_i)^2].
\label{15}
\end{equation}
The cutoff terms drop out.

If $s$ is an {\it even} integer, we obtain by an analogous argument
\begin{equation}
E=\frac{\pi s(2s+1)}{6L}-\frac{\pi(s+1)}{8L}\sum_{i=1}^s[\beta_i^2+(2-\beta_i)^2],
\label{16}
\end{equation}
where now each $\beta_i$ lies in the interval $0 < \beta_i \leq 1$.
\subsection{Contour Integration Method}
This is a very powerful method. In the context of Casimir calculations it dates back to van Kampen et al. [12]. More recently, it was applied to Casimir calculations on the spherical ball, in Refs. [13, 14]. The method was first applied to the composite string system in Ref. [3]. The starting point is the so-called argument principle, which states that any meromorphic function  $g(\omega)$  satisfies the relation
\begin{equation}
\frac{1}{2\pi i}\oint \omega \frac{d}{d\omega}\ln g(\omega)=\sum\omega_0-\sum\omega_{\infty},
\label{17}
\end{equation}
where $\omega_0$ are the zeros and $\omega_{\infty}$ are the poles of  $g(\omega)$ inside the integration contour. The contour is chosen to be a semicircle of large radius $R$ in the right half complex $\omega$ plane, closed by a straight line from $\omega=iR$ to $\omega=-iR$. The great advantage of the method - in contradistinction to the previous cutoff method - is that the {\it multiplicity} of the zeros (there are no poles in the present case) are automatically taken care of.

We make the following ansatz for $g(\omega)$:
\begin{equation}
g(\omega)=\frac{F(x)\sin^2[(s+1)\omega L_I/2]+\sin(\omega L_I)\sin(s\omega L_I)}{F(x)+1}.
\label{18}
\end{equation}
This means that $g(\omega)$ is chosen to be the expression to the left in (4), multiplied by $[F(x)+1]^{-1}$. This choice is convenient, since it allows us to perform partial integrations in the energy integral without encountering any divergences in the boundary terms when $R \rightarrow \infty$. The final result becomes $(\omega=i\xi)$
\begin{equation}
E=\frac{1}{2\pi}\int_0^{\infty}\ln \left|\frac{F(x)+\frac{\sinh \xi L_I \sinh s\xi L_I}
{\sinh^2[(s+1)\xi L_I/2]}}{F(x)+1} \right| d\xi.
\label{19}
\end{equation}
This zero-temperature result is very general; it holds for any value of $s$, not only for integers $s$ as considered in the previous subsection. Since (19) is invariant under the interchange $s \rightarrow 1/s$, it follows that $s$ can be restricted to the interval $s \geq 1$ without any loss of generality. If $x\rightarrow 0$, we recover the simple formula (10).

Another advantage of the contour integration method is that the zero-temperature result can easily be generalized to the case of finite temperatures. The integration over continuous imaginary frequencies $\xi$ then has to be replaced by a sum over discrete Matsubara frequencies $\xi_n=2\pi nk_BT,~~n=0, 1, 2,...$ We get
\begin{equation}
E(T)=k_BT{\sum_{n=0}^{\infty}}'\ln \left|\frac{F(x)+\frac{\sinh \xi_nL_I\sinh s\xi_n L_I}
{\sinh^2[(s+1)\xi_nL_I/2]}}{F(x)+1} \right|,
\label{20}
\end{equation} 
valid for any temperature $T$. The prime on the summation sign means that the $n=0$ term is taken with half weight.
\subsection{$\zeta-$ Function Method}
This elegant regularization method has proved to be most useful in many cases. General treatises on it can be found in Ref.[15], and also in Elizalde's book listed in [3]. The first application to the composite string was made by Li et al. [2]. The appropriate $\zeta-$function to be used in this case is not the Riemann function $\zeta(s)$, but instead the Hurwitz function $\zeta(s,a)$, the latter being originally defined as
\begin{equation}
\zeta(s,a)=\sum_{n=0}^{\infty}(n+a)^{-s}~~~~(0<a<1,~~~{\rm Re}~s>1).
\label{21}
\end{equation}
For practical purposes on needs only the property
\begin{equation}
\zeta(-1,a)=-\frac{1}{2}(a^2-a+\frac{1}{6})
\label{22}
\end{equation}
of the analytically continued Hurwitz function.

The $\zeta-$function method has one important property in common with the cutoff method: the eigenvalue spectrum must be determined explicitly. Consider the uniform string first: in this case the Riemann function is adequate, giving the zero-point energy
\begin{equation}
E_{uniform}=\frac{2\pi}{L}\zeta(-1)=-\frac{\pi}{6L},
\label{23}
\end{equation}
in agreement with the finite part of (7). Consider next the composite string, assuming $s$ to be an odd integer: by inserting the degenerate branch eigenvalue spectrum (11) we have
\begin{equation}
E({\rm degenerate~~branch})=-\frac{\pi}{12L_I}.
\label{24}
\end{equation}
Using the generic form (12) for the double branches we obtain analogously
\begin{equation}
E({\rm double~~branch}) = \frac{\pi}{2L_I}[\zeta(-1,\beta)+\zeta(-1,1-\beta)]  = \frac{\pi}{6L_I}-\frac{\pi}{4L_I}[\beta^2+(1-\beta)^2].
\label{25}
\end{equation}
Summing (25) over the $\frac{1}{2}(s-1)$ double branches, and adding (24), we obtain the composite string's zero-point energy
\begin{equation}
E_{I+II}=\frac{\pi (s-2)}{12 L_I}-\frac{\pi}{4 L_I}\sum_{i=1}^{(s-1)/2}[\beta_i^2+(1-\beta_i)^2].
\label{26}
\end{equation}
Now subtracting off (23), we obtain the same expression for the Casimir energy $E$ as in Eq.(15).

The case of even integers $s$ is treated analogously. The $\zeta-$function method is somewhat easier to implement than the cutoff method.

\section{$2N-$ Piece String}
\subsection{Recursion Equation and Casimir Energy}
In the same way one can consider the Casimir theory of a string of length $L$ divided into three pieces, all of the same length.  The theory for this case has been given in Refs.[5] and [8]. Here, we shall consider instead a string divided into $2N$ pieces of equal length, of alternating type $I$/type $II$ material. The string is relativistic, in the same sense as before. The basic formalism for arbitrary integers $N$ was set up in Ref.[4], but the Casimir energy was there calculated in full only for the case of $N=2$. A full calculation was worked out in Ref.[7]; cf. also Ref.[8]. A key point in [7] was the derivation of a new recursion formula, which is applicable for general integers $N$.

We introduce two new symbols, $p_N$ and $\alpha$:
\begin{equation}
p_N=\omega L/N,~~~~~\alpha=(1-x)/(1+x).
\label{27}
\end{equation}
The eigenfrequencies are determined from
\begin{equation}
{\rm Det}[{\bf M}_{2N}(x,p_N)-{\bf 1}]=0.
\label{28}
\end{equation}
Here it is convenient to scale the resultant matrix $ {\bf M}_{2N}$ as
\begin{equation}
{\mathbf M}_{2N}(x,p_N)=\left[ \frac{(1+x)^2}{4x}\right]^N{\mathbf m}_{2N}(\alpha, p_N),
\label{29}
\end{equation}
and to write $ {\bf m}_{2N} $ as a product of component matrices:
\begin{equation}
{\bf m}_{2N}(\alpha, p_N)=\prod_{j=1}^{2N}{\bf m}^{(j)}(\alpha, p_N),
\label{30}
\end{equation}
with
\begin{equation}
{\mathbf m}^{(j)}(\alpha,p_N)=\left( \begin{array}{ll}
                                   1, & \mp \alpha e^{-ijp_N}\\
                                   \mp \alpha e^{ijp_N}, & 1
                                     \end{array} \right)
\label{31}
\end{equation}
for $j=1, 2,...(2N-1)$. The sign convention is to use +/- for even/odd $j$. At the last junction, for $j=2N$, the component matrix has a particular form (given an extra prime for clarity):
\begin{equation}
{\mathbf m}'^{2N}(\alpha,p_N)=\left( \begin{array}{ll}
                                   e^{-iN p_N}, & \alpha e^{-i N p_N}\\
                                   \alpha e^{iN p_N}, & e^{iN p_N}
                                    \end{array} \right) .
\label{32}
\end{equation}
Now the recursion formula alluded to above can be stated:
\begin{equation}
{\mathbf m}_{2N}(\alpha, p_N)={\mathbf \Lambda}^N(\alpha, p_N),
\label{33}
\end{equation}
where $ \mathbf \Lambda $ is the matrix
\begin{equation}
{\mathbf \Lambda}(\alpha,p)=\left( \begin{array}{ll}
                                 a & b\\
                                 b^* & a^*
                                 \end{array} \right) ,
\label{34}
\end{equation}
with
\begin{equation}
a=e^{-ip}-\alpha^2, ~~~~~b=\alpha (e^{-ip}-1).
\label{35}
\end{equation}
The obvious way to proceed is now to calculate the eigenvalues of $\mathbf \Lambda$, and express the elements of $ {\mathbf M}_{2N}$ as powers of these. More details can be found in [7].

Consider next the Casimir energy. The most powerful regularization method, as above, is the contour regularization method. Using it we obtain, for arbitrary $x$ and arbitrary integers $N$, at zero temperature,
\begin{equation}
E_N(x)=\frac{N}{2\pi L}\int_0^{\infty}\ln \left| \frac{2(1-\alpha^2)^N-[\lambda_+^N(iq)+\lambda_-^N(iq)]}
{4 \sinh^2 (Nq/2)} \right| dq.
\label{36}
\end{equation}
Here $\lambda_{\pm}$ are eigenvalues of $\mathbf \Lambda$, for imaginary arguments $iq$, of the dispersion equation. Explicitly,
\begin{equation}
\lambda_{\pm}(iq)=\cosh q-\alpha^2 \pm [(\cosh q-\alpha^2)^2-(1-\alpha^2)^2]^\frac{1}{2}.
\label{37}
\end{equation}
Evaluation of the integral shows that $E_N(x)$ is negative, and the more so the larger is $N$. A string can thus in principle always diminish its zero-point energy by dividing itself into a larger number of pieces of alternating type I/II material.

In the limiting case of $x \rightarrow 0$ the integral can be solved exactly:
\begin{equation}
E_N(0)=-\frac{\pi}{6 L}(N^2-1).
\label{38}
\end{equation}
The generalization of (36) to the case of finite temperatures is easily achieved following the same method as above. 

As an alternative method, on can insted of contour integration make use of the $\zeta-$ function method; one then has to determine the spectrum explicitly and thereafter put in the degeneracies by hand. The latter mehod is therefore most suitable for low $N$.
\subsection{Scaling Invariance}
A rather unexpected scaling invariance property of the Casimir energy becomes apparent if we examine the behaviour of the function $f_N(x)$ defined by
\begin{equation}
f_N(x)=\frac{E_N(x)}{E_N(0)}.
\label{39}
\end{equation}
This function generally has a value that lies between zero and one. If we calculate $E_N(x)$ (usually numerically) versus $x$  for some fixed value of $N$, we find that the resulting curve for  $f_N(x)$ is practically the {\em same}, irrespective of the value of $N$, as long as $N \geq 2$. (The case $N=1 $ is exceptional, since $E_1(x)=0$.)
Numerical trials show that the simple analytical form 
\begin{equation}
f_N(x) \rightarrow f(x)= (1-\sqrt x )^{5/2}
\label{40}
\end{equation}
is a useful approximation, in particular in the region $0<x<0.45$. The special form of Eq.(40) appears to be related to the applicability of the so-called Puisseux mathematical series to the present problem. These topics are discussed in more detail in Ref.[16]. A profound physical interpretation of this striking scaling invariance of the Casimir energy has so far not been obtained.

\section{Thermodynamic Properties}
Reference [10] contains a calculation of the thermodynamic free energy $F$ for a gas whose particles are the quantum excitations of a two-piece string. A flat, $D-$dimensional, spacetime is assumed. The length ratio $s$ is taken to be an integer.  Our calculation is carried out in full only in the limiting case when $x \rightarrow 0$. From the dispersion relation we obtain two sequences of modes: one sequence  belonging to the first branch, called $\omega_n(s)$, and one belonging to the second branch, called $\omega_n(s^{-1})$.  Explicitly,
\begin{equation}
\omega_n(s)=(1+s)n,~~~~\omega_n(s^{-1})=(1+s^{-1})n,
\label{41}
\end{equation}
where $n= \pm 1, \pm 2, \pm3,...$.
 
Let now $X^\mu(\sigma,\tau)$ ($\mu=0, 1, 2,.. (D-1)$) specify the coordinates on the world sheet. For each of the branches we can write the general expression for $X^\mu$ in the form
\begin{equation}
X^\mu=x^\mu+\frac{p^\mu \tau}{\pi \overline {T}(s)}+\theta(L_I-\sigma)X_I^\mu+\theta (\sigma-L_I)X_{II}^\mu,
\label{42}
\end{equation}
where $x^\mu$ is the centre-of-mass position and $p^\mu$ is the total momentum of the string. $\overline{T}(s) = T_{II}s/(1+s)$ is the mean tension, and $\theta$ is the step function, $\theta(x>0)=1, ~~\theta(x<0)=0$. 

We consider henceforth the first branch only. Then, in region $I$,
\begin{equation}
X_I^\mu=\frac{i}{2 \sqrt{\pi T_I}}\sum_{n \neq 0}\frac{1}{n}\left[\alpha_n^\mu(s)e^{i(1+s)n(\sigma-\tau)}
+\tilde{\alpha}_n^\mu(s)e^{-i(1+s)n(\sigma+\tau)}\right],
\label{43}
\end{equation}
where the $\alpha_n,~\tilde{\alpha}_n$ are oscillator coordinates of the right- and left- moving waves. The analogous expansion for the first branch in region $II$ can be written
\begin{equation}
X_{II}^\mu=\frac{i}{2\sqrt{\pi T_I}}\sum_{n\neq 0}\frac{1}{n}\gamma_n^\mu(s)e^{-i(1+s)n\tau}
\cos[(1+s)n\sigma],
\label{44}
\end{equation}
where $\gamma_n$ means
\begin{equation}
\gamma_n^\mu(s)=\alpha_n^\mu(s)+\tilde{\alpha}_n^\mu(s), ~~~~n\neq 0.
\label{45}
\end{equation}
The oscillations in region $II$ are thus standing waves; this being a direct consequence of the junction conditions in the limit $x \rightarrow 0$.
  
In the quantum theory of the system the starting point is the following expression for the free energy $F$, at finite temperature $T$, of free fields of mass $F$ in $D$ dimensions:
\begin{equation}
\beta F= -\ln Z = \frac{1}{2}\beta\sum_{-\infty}^{\infty}\omega_n-\beta\sum_{m=1}^{\infty}
\int_0^{\infty}\frac{du}{u}(2\pi u)^{-D/2}\exp\left( -\frac{M^2u}{2}-\frac{m^2\beta^2}{2u}\right),
\label{46}
\end{equation}
where $\beta=1/k_BT$, $Z$ being the partition function. The equal-time commutation rules are taken to be
\begin{equation}
T_I[\dot{X}^\mu(\sigma,\tau), X^\nu(\sigma', \tau)]=-i\delta(\sigma-\sigma')\eta^{\mu\nu}
\label{47}
\end{equation}
in region $I$, and a similar relation with $T_I \rightarrow T_{II}$ in region $II$. Here $\eta^{\mu\nu}$ is the $D-$dimensional Minkowski metric. The other commutation relations vanish. By introducing annihilation and creation operators $a_n^\mu, ~c_n^\mu$ (and their Hermitean conjugates) for the first branch in the following way:
\begin{equation}
\alpha_n^\mu(s)=\sqrt{n}\, a_n^\mu(s),~~~~~\gamma_n^\mu(s)=\sqrt{4nx}\, c_n^\mu(s),
\label{48}
\end{equation}
we find for $n\geq1$ the standard form
\begin{equation}
[a_n^\mu(s),{a_m^\nu}^{\dagger}(s)]=\delta_{nm}\eta^{\mu \nu},~~~~
[c_n^\mu(s),{c_m^\nu}^{\dagger}(s)]=\delta_{nm}\eta^{\mu\nu}.
\label{49}
\end{equation}
We can herefrom write the total Hamiltonian $H$ as a sum of two parts, $H=H_I+H_{II}$, where
\begin{equation}
H_I=-\frac{M^2x}{2st(s)}+\frac{1}{2}\sum_{n=1}^{\infty}\omega_n(s)
[a_n^{\dagger}(s)\cdot a_n(s)+\tilde{a}_n^{\dagger}(s)\cdot\tilde{a}_n(s)],
\label{50}
\end{equation}
\begin{equation}
H_{II}=-\frac{M^2}{2t(s)}+s\sum_{n=1}^{\infty}\omega_n(s)c_n^{\dagger}(s)\cdot c_n(s),
\label{51}
\end{equation}
with $t(s)=\pi \overline{T}(s)$.  The basic condition imposed on the system is that $H=0$ for the physical states. From this condition one can calculate the mass $M$ via the relation $M^2=-p^\mu p_\mu$, similarly as in the case of the uniform string [17]. Inserting the expression for $M^2$ into Eq.(46), and using that in the limit $x\rightarrow 0$ one has
\begin{equation}
\frac{1}{2}\sum_{-\infty}^\infty \omega_n\rightarrow -\frac{1}{24}\left(s+\frac{1}{s}-2\right)
\label{52}
\end{equation}
(we put $L=\pi$), we can calculate $F$.  Since the expression for $F$ is somewhat complicated [10], it will not be given here. Let us instead consider one of the consequences from $F$, namely the expression for the Hagedorn temperature $T_c$. This is the temperature above which the free energy is ultraviolet divergent. We get \footnote{This differs from the expression for $T_c$ obtained in [8].}
\begin{equation}
k_B T_c=\frac{s}{4}\sqrt{\frac{T_{II}}{\pi(1+s)}}.
\label{53}
\end{equation}
In particular, if one of the pieces of the string is much shorter than the other ($s \rightarrow \infty $, implying that a point "mass" is sitting on an otherwise uniform string), letting $T_{II}$ be a fixed finite quantity, we see that $T_c$ becomes proportional to $\sqrt{s}$. Thus $T_c \rightarrow \infty$ when $s\rightarrow \infty$, implying that the free energy is always  ultraviolet finite. 

\end{document}